\documentclass{cidr-modefied}
\usepackage[utf8]{inputenc}
\usepackage{booktabs}
\usepackage{hyperref}
\usepackage{url}
\usepackage{authblk}

\hypersetup{
colorlinks,
linkcolor=blue,
urlcolor=blue,
}
\newcommand{\bestName}{{\sf QueenBee}}

\begin{document}
\title{Decentralized Search on Decentralized Web}
\author[1]{Ziliang Lai}
\author[1]{Chris Liu}
\author[1]{Eric Lo}
\author[2]{Ben Kao}
\author[2]{Siu-Ming Yiu}
\affil[1]{Chinese University of Hong Kong (CUHK)}
\affil[2]{The University of Hong Kong (HKU)}

\maketitle

Decentralized Web, or DWeb, is envisioned as a promising future of the Web \cite{kahle}. 
Being decentralized, there are no dedicated web servers in DWeb;
Devices that retrieve web contents also serve their cached data to peer devices with straight privacy-preserving mechanisms.
The fact that contents in DWeb are distributed, replicated, and decentralized lead to a number of key advantages over the conventional web. These include better resiliency against network partitioning and distributed-denial-of-service attacks (DDoS)~\cite{DoS}, and
better browsing experiences in terms of shorter latency and higher throughput.
Moreover, DWeb provides tamper-proof contents because each content piece is uniquely identified by a cryptographic hash.
A DWeb prototype, which hosts a Wikipedia snapshot, can be found 
\href{https://ipfs.io/ipfs/QmXoypizjW3WknFiJnKLwHCnL72vedxjQkDDP1mXWo6uco/wiki/index.html}{here}.
DWeb also clicks well with future Internet architectures, such as {\it Named Data Networking} (NDN)~\cite{NDN}.


Search engines have been an inseparable element of the Web. 
Contemporary (``Web 2.0") search engines, however, provide centralized services. They are thus  subject to DDoS attacks~\cite{DoS}, insider threat \cite{insider}, and ethical issues like search bias \cite{bias} and censorship \cite{censorship}.
As the web moves from being centralized to being decentralized, search engines ought to follow.
We propose \bestName, a decentralized search engine for DWeb. 
\bestName\ is so named because worker bees and honeycomb are a common metaphor for distributed architectures, with the queen being the one that holds the colony together.

\begin{figure}
\includegraphics[width=\columnwidth]{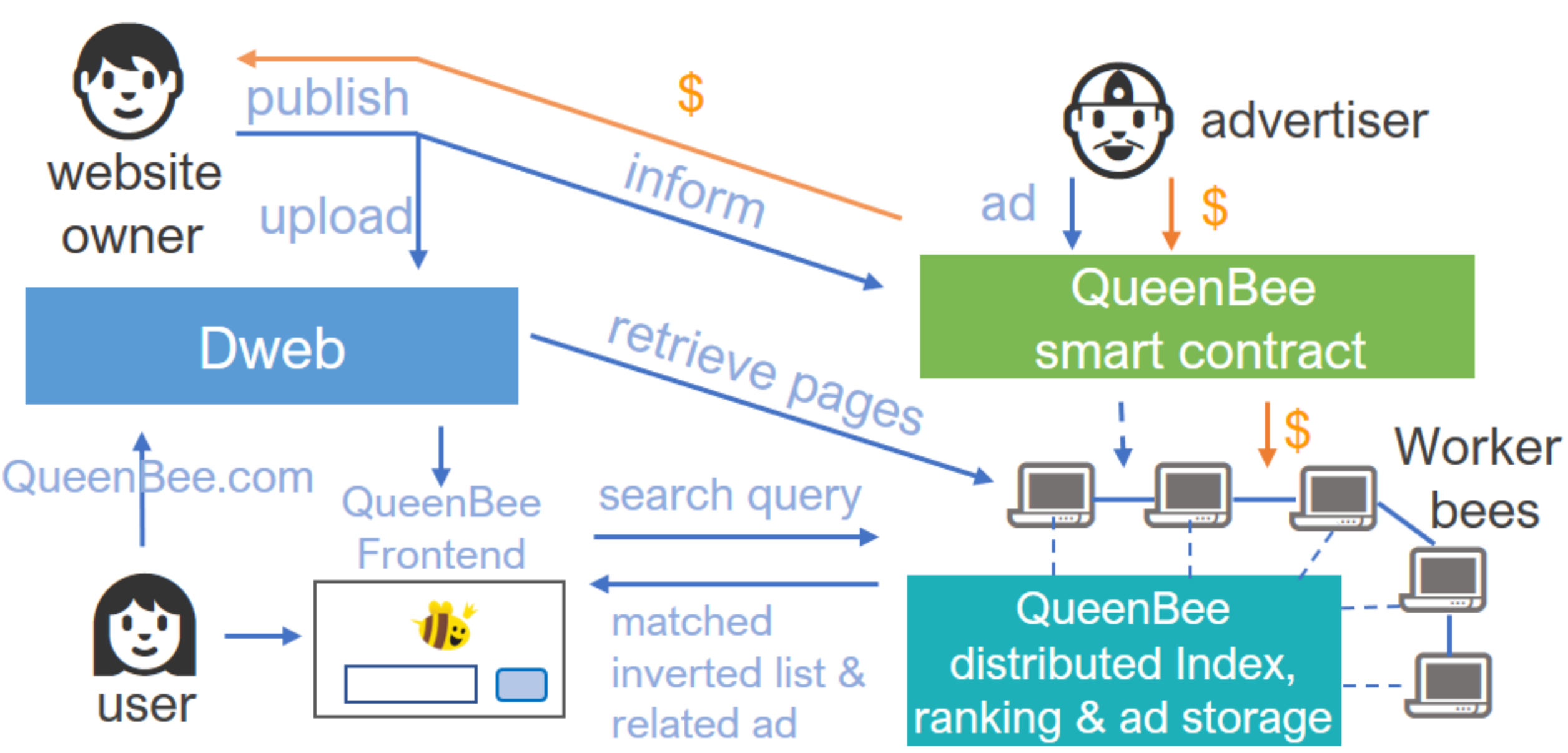}
\caption{\bestName~and the Dweb}
\label{framework}
\end{figure}

\bestName~aims to revolutionize the search engine business model by offering incentives to both  content providers and peers that participate in \bestName's page indexing and ranking operations. 
Figure \ref{framework} shows our vision of \bestName, whose 
core business operations are autonomously and securely governed by \emph{smart contracts} deployed on a cryptocurrency blockchain like Ethereum \cite{ethereum}.


\bestName~advocates \emph{no-crawling}, because crawling inevitably reduces the freshness of the search results.  Instead, \bestName~incentivizes content creators to 
{\it publish} (create or update) their contents via \bestName's smart contract
to gain ``{honey}''
in the form of a cryptocurrency. 
{Honey} is also rewarded to {\sf worker bees} -- peers that help 
update the index and compute the page ranks, which are hosted in a decentralized storage (e.g., IPFS \cite{ipfs}).
Users submit their keyword queries via \bestName's HTML+Javascript frontend\footnote{Readers can experience the QueenBee prototype \href{http://ipfs.io//ipns/QmZkAQebh5CWvsayZNneejVEPVpk1jQPnt1pUphTJQ8g6M/queenbee.html}{here}.} on the DWeb.  The frontend is also responsible for composing the search results by intersecting the matched inverted lists, ranking the results, and displaying relevant ads. 

\bestName~is a \emph{decentralized organization} -- 
advertisers directly make advertisements through our smart contract and \emph{the ad revenue is shared among the content creators and worker bees}.   
\bestName's decentralized nature rids itself of problematic issues (e.g., search bias) found in centralized search engines. 
To our best knowledge,  \bestName~is the world first initiative to build a decentralized search engine on the DWeb. 
Existing P2P search engines (e.g., YaCy \cite{yacy}) only work on Web 2.0, without an incentive scheme or a security incentive that guard against practical attacks.

Besides performance issues, \bestName\ will face many new and interesting research challenges. We briefly discuss two of them. (I) \emph{A fair incentive scheme for all stakeholders}: For example, while allowing any content provider to use our service, we need to reward those whose websites are popular. A simple way is to give the providers for which the page ranks of their websites exceed a certain threshold some \bestName's honey. For advertisers, we also need a fair scheme to charge them (e.g., they pay by the number of clicks on the ad). In general, a sensible scheme is needed to maintain the ecosystem of \bestName. (II) \emph{New attacks}: this new model of decentralized search engine may induce new attacks. For examples, 
 an attack from colluded worker bees that aim at manipulating \bestName's indexes or page ranking data maliciously (\emph{collusion attack}); as popular webpages will gain \bestName's honey, \emph{scrapper site attack} may exist that tries to mirror popular websites for \bestName's honey.

\vspace{-0.2cm}
\setcounter{secnumdepth}{0}
\bibliographystyle{abbrv}

\bibliography{bibliography}

\end{document}